\documentclass{iau}
\usepackage{graphicx}

\title[The Densities of Planets and the Masses of Host Stars] 
{The Densities of Planets and Masses of Host Stars}

\author[Eike W. Guenther \& CoRoT-Team]   
{Eike W. Guenther$^1$
 \and CoRoT-Team}

\affiliation{$^1$ Thueringer Landessternwarte Tautenburg,
D-07778 Tautenburg, Germany \\ email: {\tt guenther@tls-tautenburg.de} \\[\affilskip]}
\pubyear{2013}
\volume{293}  
\pagerange{100--106}
\jname{Formation, Detection, and Characterization of Extrasolar Habitable Planets}
\editors{N. Haghighipour}
\begin{document}

\maketitle

\begin{abstract}

Studies of transiting extra-solar planets are of key importance for
understanding the nature of planets outside our Solar System, because
their densities can be determined, constraining of what the planets are
made of.  Using the data obtained by the CoRoT space telescope we
study the relation between the density of planets, their mass, and the
mass of their host stars. Although planets of the same mass can have
different densities, we still find some trends.  Planets with more
than 1000 $M_{Earth}$ ($\sim$ 3 $\rm M_{Jup}$) have densities larger
than 6 $\rm g\,cm^{-3}$, and are preferentially found in stars that
are more massive than the Sun. All known planets in the mass-range
between 15 and 600 $M_{Earth}$ ($\sim$ 0.05 to 2 $M_{Jup}$) have
densities of less than 3 $\rm g\,cm^{-3}$.  When going further down in
the mass of the planets, the density increases steadily, and there is
no sudden transition from gaseous to rocky planets. Based on the
current sample, we do not find any difference for planets at different
galactocentric distances.

\end{abstract}

\firstsection 

\section{Introduction}

Studies of transiting extra-solar planets are of key importance for
understanding the nature of planets outside our Solar System, because
their mass, diameter, and hence their bulk density can be derived.
The density is a key parameter, because it gives us information on the
structure and composition of the planets.  Space-telescopes like CoRoT
and Kepler are particularly suitable for such studies, because they
allow us to study large and small planets orbiting different types of
stars.

In our Solar System we have basically two species of planets: gas or
ice giants which have masses larger than 15 $M_{Earth}$ ($\sim$ 0.05
$\rm M_{Jup}$) and densities between 0.7 to 1.6 $\rm gcm^{-3}$, and
rocky planets (Mercury, Venus, Earth and Mars) which have densities
from 3.7 to 5.5 $\rm g\,cm^{-3}$.  It is thus natural to expect that
planets outside the Solar System should have the same properties:
Planets with more than 15 $\rm M_{Earth}$ should have a low density,
and planets with the mass of the Earth should have a high density. The
transition between these two types of planets is expected to be
somewhere between 1 and 15 $\rm M_{Earth}$. In here we compare the
results obtained with CoRoT with expectations that small planets
should have a high density, an large planets a small density. We also
discuss whether there is a relation between the properties of planets
and the masses and galactocentric distances of their host stars.

\section{The CoRoT mission}

CoRoT (COnvection ROtation \& planetary Transits) is a space mission
that is especially designed to monitor the brightness changes of
stars. The objectives of the mission are the study of stellar
oscillations and the detection of extrasolar planets. The satellite
was launched on December 27 2006. In the exoplanet part of the
mission, CoRoT monitors typically 6000 to 12000 in per field. Up to
now 24 fields have been observed. The fields are located in two
opposite directions in the sky: The so-called "galactic center eye"
(galactic longitude: $\sim 213\pm5^o$), and the so-called "galactic
anti-center eye" (galactic longitude: $\sim 33\pm5^o$).  Given that
CoRoT observes stars between 11 and 16 mag, the F and G-stars that
CoRoT observes have galactocentric distances between 7 to 9 kpc which
corresponds to the galactic habitable zone (Lineweaver et al.
\cite{lineweaver04}).

CoRoT has detected a transiting planet with a relative transit depth
of $3.35\,10^{-4}$ orbiting a star of 11.7 mag (Leger et
al. \cite{leger09}). Thus, CoRoT is sensitive enough to detect planets
with less than two Earth-radii for solar-like stars, and Jupiter-sized
planets of B4V-stars. From the CoRoT-data, we obtain a frequency of
transiting hot Jupiters orbiting solar-like stars of $\rm
0.05_{-0.01}^{+0.02}\%$. This corresponds to a frequency of hot
Jupiters of $0.4_{-0.1}^{+0.2}\%$, which agrees well to the rate found
in radial-velocity surveys (Cumming et al. \cite{cumming08}; Naef et
al. \cite{naef05}).  This means that CoRoT has detected all transiting
hot Jupiters in the fields observed.

\section{The density of planets and the mass of the host star}

\begin{figure}
\begin{center}
 \includegraphics[width=0.7\textwidth, angle=-90]{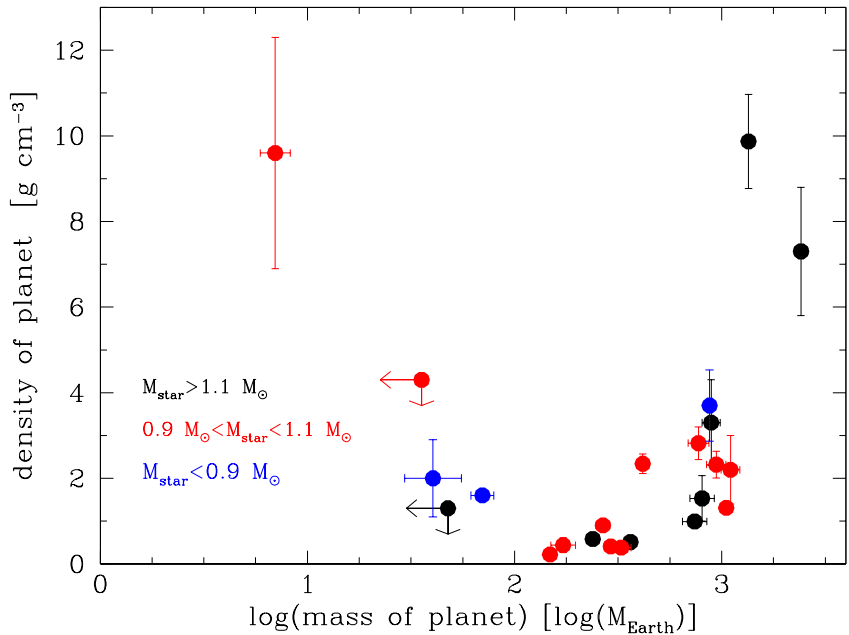}
\vspace*{0.0 cm}                                                                                                                           
\caption{Mass-density diagram for planets discovered by CoRoT.
Errors denote upper limits.}
   \label{fig1}
\end{center}
\end{figure}

\begin{figure}
\begin{center}
 \includegraphics[width=0.7\textwidth, angle=-90]{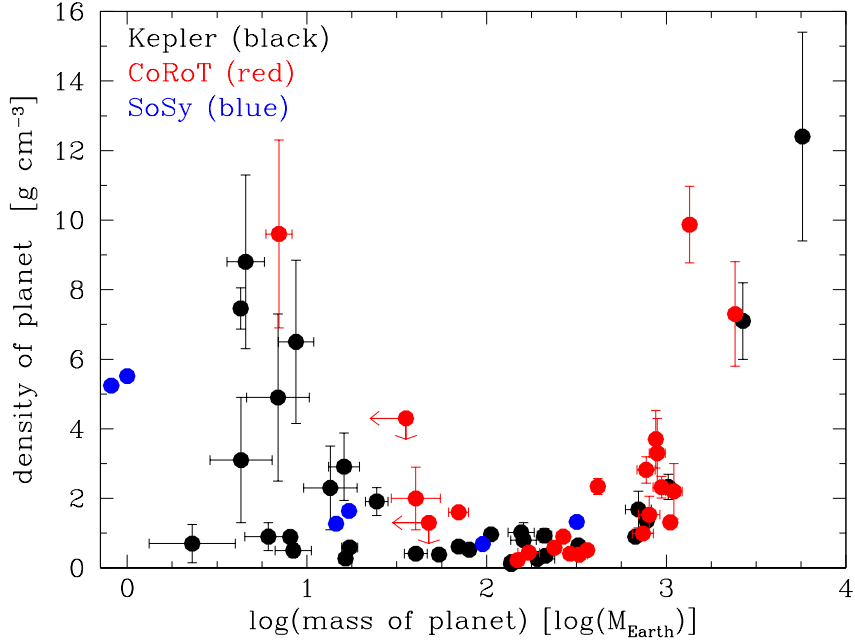} 
\vspace*{0.0 cm}

\caption{Mass-density diagram for the planets discovered by CoRoT
  (red) and Kepler (black) compared to objects in the solar-system
  (blue). Note that 13 (or 25) $M_{Jup}$ corresponding to 
  $\rm log (M_{Earth})=3.6 (3.9)$ are usually referred as
  the boarder line between planets and brown dwarfs 
  (Schneider et al. \cite{schneider11}).}
   \label{fig2}
\end{center}
\end{figure}

An important discovery made by CoRoT was CoRoT-7b, a planet with a
mass of $\rm M_p=7.42\pm1.21$ $\rm M_{Earth}$, and a density of $\rm
\rho=10.4\pm1.8$ $\rm g\,cm^{-3}$ (detection: L{\'e}ger et
al. \cite{leger09}; Queloz et al. \cite{queloz09}; new mass
determination: Hatzes et al.  \cite{hatzes11}). This planet thus was
the first rocky planet found outside the solar-system.  Kepler-10b is
another object like this. It has a mass of $\rm
M_p=4.56^{+1.17}_{-1.29}$ $\rm M_{Earth}$, and a density of
$\rho=8.8^{+2.1}_{-2.9}$ $\rm g\,cm^{-3}$ (Batalha et al.
\cite{batalha11}). These two planets thus can be nicknamed
``super-Earths'', planets more massive than the Earth but with a
density that is consistent with an Earth-like composition (Valencia et
al. \cite{valencia10}). The only other known low-mass planets ($\rm
M_p\leq 10 M_{Earth}$) with densities larger than 4 $\rm g\,cm^{-3}$
are Kepler-18b ($\rm M_p= 6.9\pm3.4 M_{Earth}$, $\rm \rho = 4.9\pm2.4
g\,cm^{-3}$), Kepler-20b ($\rm M_p= 8.7^{+2.1}_{-2.2} M_{Earth}$, $\rm
\rho = 6.5^{+2.0}_{-2.7} g\,cm^{-3}$), and Kepler-36b ($\rm M_p=
4.45^{+0.33}_{-0.27} M_{Earth}$, $\rm \rho = 7.46^{+0.07}_{-0.05}
g\,cm^{-3}$).  All other low-mass planets have densities lower than 4
$\rm g\,cm^{-3}$.

Fig.\,\ref{fig1} shows the mass-density diagram for the 25
CoRoT-planets.  Planets orbiting stars with masses between 0.9 and 1.1
$M_{\odot}$ are marked with red dots in Fig.\,\ref{fig1}. Planets
orbiting stars of higher mass are marked as black dots, and planets
orbiting stars of lower mass as blue dots.  The figure shows that
planets of the same mass can have quite different densities.

Another interesting aspect is that not only low-mass planets, like
CoRoT-7b, can have a high density but also planets of several $\rm
M_{Jup}$.  Particularly interesting is CoRoT-20b, a planet with a
density of $8.87\pm1.10$ $\rm g\,cm^{-3}$ , and a mass of
$4.24\pm0.23$ $\rm M_{Jup}$ ($1348\pm73$ $\rm M_{Earth}$).  This high
density implies that the core amounts to between 50 to 75\% of the
total mass of this planet (Deleuil et al. \cite{deleuil11}).  Another
object like this is CoRoT-14b, which has a density of $7.3\pm1.5$ $\rm
g\,cm^{-3}$, and a mass of $7.6\pm0.6$ $M_{Jup}$ ($2420\pm190$ $\rm
M_{Earth}$)(Tingley et al.  \cite{tingley11}). Interestingly, both
planets are the most massive planets found by CoRoT.  Is this just a
coincidence, or do all massive planets have such high
densities?  Fig.\,\ref{fig2} shows the same diagram as in
Fig.\,\ref{fig1} but also including the 34 planets discovered by
Kepler with measured densities, and the planets of our solar-system.
The Kepler results agree very well with the CoRoT results.  Also the
planets with more than 1000 $M_{Earth}$ found by Kepler have densities
larger than 6 $\rm g\,cm^{-3}$.  It thus seems that massive planets in
general have a high density. Contrary to this, no planets with a
density higher than 3 $\rm g\,cm^{-3}$ has been found in the whole
mass-range between 15 and 600 $\rm M_{Earth}$.

\section{The mass of the planets and the mass of the host star}

\begin{figure}
\begin{center}
 \includegraphics[width=0.7\textwidth, angle=-90]{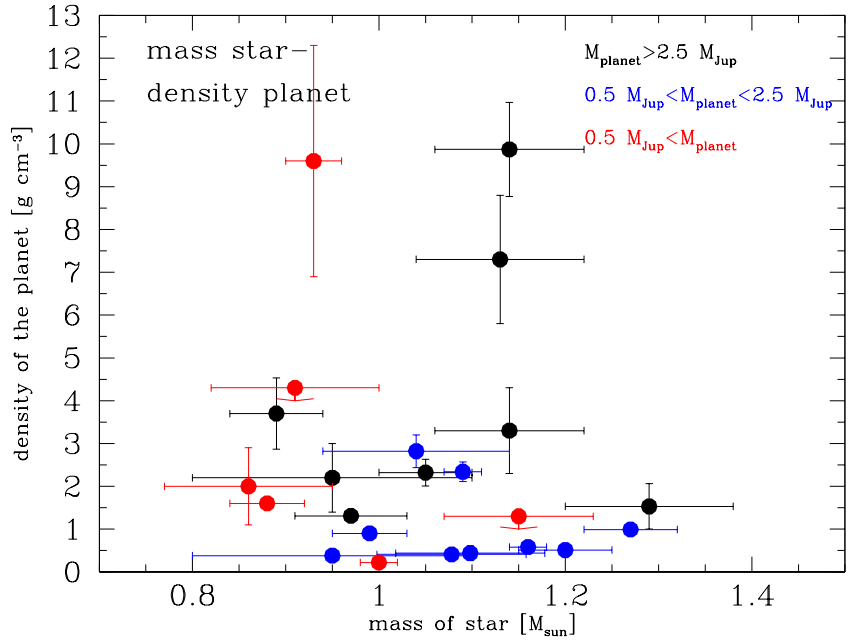}
\vspace*{0.0 cm}                                                                                                                       
\caption{Relation between the density of planets and the masses of the
  host stars.}
   \label{fig3}
\end{center}
\end{figure}

Another interesting aspect is that the two most massive planets that
CoRoT has found are orbiting stars that are more massive than the Sun.
Since the two brown dwarf companions found by CoRoT, CoRoT-3b (Deleuil
\cite{deleuil08}) and CoRoT-15b (Bouchy (\cite{bouchy11}), are also
orbiting stars more massive than the Sun, there seems to be a trend
that massive planets and brown dwarfs are preferentially found in
higher mass stars. Does this mean that there is a close relation
between the mass or density of a planet and the mass of its host star?
Fig.\,\ref{fig3} shows the relation between the density of planets and
the masses of the host stars. Planets of different mass-intervals are
marked with different colours.  Apart from the fact that massive
substellar companions are preferentially found in more massive stars,
there is no further connection between the mass, or density of planets
and the masses of host stars (see also Fig.\,\ref{fig5}).

\begin{figure}
\begin{center}
 \includegraphics[width=0.7\textwidth, angle=-90]{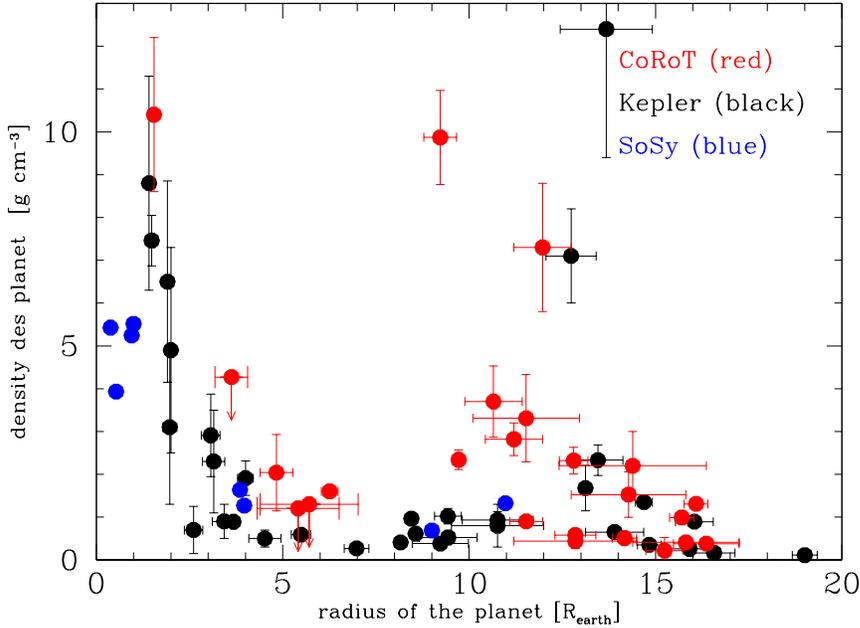}
\vspace*{0.0 cm}                                                                                                                          
 \caption{Relation between the radius and the density of planets.}
   \label{fig4}
\end{center}
\end{figure}

\section{The frequency of planets and the mass of the host star}

So far CoRoT has measured the mass and radii of 25 planets, of which 8
are orbiting F-stars, 13 are orbiting G-stars, 4 are orbiting K-stars.
Does this mean that planets are more common in G-type stars than in
other types of stars? Certainly not, we have to know how many F, G,
and K-stars the sample contains.  In order to characterize the sample
that CoRoT did observe, we have obtained spectra of 11466 stars in 3
of the 24 CoRoT-fields (Guenther et al. \cite{guenther12}; Sebastian
\cite{sebastian12}).  About 4\% of the stars that CoRoT observes are
main sequence B-stars, 16\% A-stars, 35\% F-stars, 15\% G-stars, 5\%
K-stars. Less than 1\% of the stars are M-dwarfs. Taking the frequency
of stars in the sample into account, we find that the frequency of
close-in massive planets around F-stars is less or equal to that of
G-stars. We thus do not find any significant increase of the frequency
of close-in planets with the mass of the host star.

\section{The radius of planets and their mass}

The first parameter that can usually be determined when a planet is
discovered in a transit search program is its radius. It would thus be
helpful if there were a relation between the radius of the planet and
its mass or density. If there were such a relation, we could
immediately focus on the most interesting objects.  Fig.\,\ref{fig4}
shows the relation between the radius and the density for planets. The
only feature that can be seen is that planets with 3 to 10 $R_{Earth}$
have densities below 3 $\rm g\,cm^{-3}$. Planets with radii larger
than 10 $R_{Earth}$, and planets with radii below 3 $R_{Earth}$ can
have a low, or a high density.  Thus, the measurement of the radius of
a planet alone does not immediately tell us what it is.

\section{Do the properties of the planets change with the 
galactocentric distance?}

\begin{figure}
\begin{center}
 \includegraphics[width=0.7\textwidth, angle=-90]{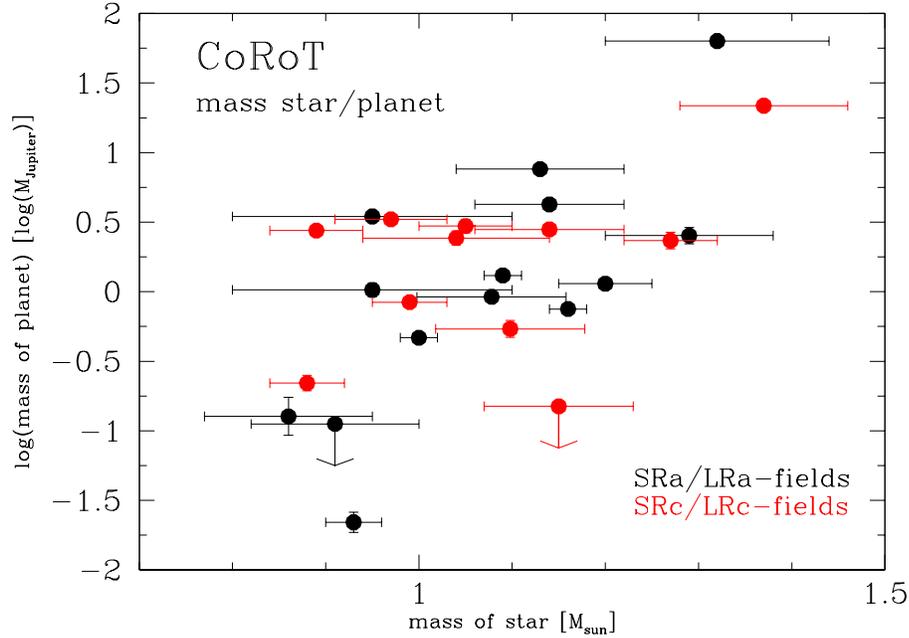}
\vspace*{0.0 cm}                                                                                                                          
 \caption{Relation between the mass of the planets and the mass of the
host stars for the two regions that CoRoT observes.}
   \label{fig5}
\end{center}
\end{figure}

CoRoT observes fields in two opposite directions in the sky.  F- and
G-stars in the so-called galactic center eye (SRc/LRc-fields) have
typical galactocentric distances of about 7 kpc, and F and G-stars in
the opposite direction (SRa/LRa-fields) about 9 kpc.  Fig.\,\ref{fig5}
shows the relation between the mass of the discovered planets and the
mass of the host stars for the two CoRoT "eyes".  We do not see any
significant difference between the planets in the two regions.

\section{Summary and conclusions}

Based on the currently available statistics we draw the following
conclusions:

   \begin{itemize}
      \item Planets of the same mass can have different densities.
      \item Planets with more than 1000 $M_{Earth}$ ($\sim$ 3
        $M_{Jup}$) have densities larger than 6 $\rm g\,cm^{-3}$. Such planets
        are preferentially found around stars that are more massive than
        the Sun.
      \item All planets in the mass-range from 15 to 600 $M_{Earth}$
        ($\sim$ 0.05 to 2 $M_{Jup}$) have densities below 3
        $\rm g\,cm^{-3}$.
       \item When going further down in mass of the planets, the 
         density increases steadily, there is no sudden transition from
         gaseous to rocky planets.
      \item Up to now we do not see any significant difference
        between the masses, or densities of the planets at different
        galactocentric distances.
      \item The frequency of close-in planets of F-stars in not higher
        than that of G-stars.
      \item The measurement of the radius of planets alone is not sufficient
        for characterizing it.
   \end{itemize}


\end{document}